**“We’ll have to see how it works”: An interview study to understand collaborative practices in interdisciplinary artificial intelligence and healthcare research**

Rafael Henkin[1]¶*, Elizabeth Remfry[1]¶, Duncan J. Reynolds[2], Megan Clinch[2], Michael R. Barnes[1]

[1] William Harvey Research Institute, Queen Mary University of London, London, United Kingdom

[2] Wolfson Institute of Population Health, Queen Mary University of London, London, United Kingdom

* Corresponding author

E-mail: r.henkin@qmul.ac.uk (RH)

¶ These authors contributed equally to this work

# Abstract

Developing artificial intelligence (AI) algorithms for healthcare is a collaborative effort, bringing data scientists, clinicians, patients and other stakeholders together. By understanding AI as ‘sociotechnical’ where the social and the technical nature of the work and the models are inseparable, we explore the AI development workflow and how stakeholders navigate the challenges and tensions of sharing and generating knowledge across disciplines. We conducted an inductive thematic analysis of 13 semi-structured interviews with participants in early stages of AI-in-healthcare research consortia in the UK. Our findings identify that participants needed to adapt both the tools used for sharing and the information communicated according to their audience, particularly when working with those with a clinical or patient perspective. We identify the novelty of participating in AI research, how AI knowledge is shared, and the inclusion of clinician and patient stakeholder perspectives as key areas within collaborative AI practices in healthcare. These findings highlight that bringing AI into the mix can introduce new obstacles to interdisciplinary work.

# Introduction

The use of artificial intelligence (AI) in healthcare is growing, from disease prediction to patient stratification and beyond (1–3). Creating AI algorithms and applying them to healthcare research requires a collaborative effort, bringing together the expertise of clinicians, data scientists and other specialists in their respective fields (4). Collaborating across disciplines has a rich history in medicine (5) and sciences (6) and is underpinned by the assumption that multiple perspectives are better at addressing complex problems (7–9). Developing and applying AI algorithms in healthcare research, and making sense of data, is a complex task as knowledge is often dispersed throughout the team and demands both domain knowledge of medicine and medical data and technical expertise of data science and AI.

In this study, we seek to understand how diverse disciplinary teams collaborate to work with data to develop AI algorithms in the early stages of research projects. We take as our starting point the position that the development of AI algorithms is socio-technical (10–12). From this perspective, we reject the idea of AI as a purely technical system, recognising that AI development always unfolds through relations with social worlds. Technology and society continuously shape one another rather than acting in a linear or unidirectional manner (13). From this standpoint, AI research and development involve not only technical choices such as model architectures or training strategies, but also social arrangements including tools, practices, institutional priorities, political interests, and cultural expectations (14). These parameters are directly influenced by the introduction of AI, due to its computational complexity, the need for large datasets in a circular manner (i.e., AI is better for large datasets than simpler methods, so AI requires large datasets when used), and its impact on disciplines due to narratives of transformative potential of AI (15).

Previous research has established important foundations for understanding collaboration in AI, data science and healthcare contexts. Work examining data science collaborations (with and without AI) in corporate and industry settings has revealed concrete challenges: communication between data scientists and domain experts is time intensive, requiring careful preparation of digital artefacts like slides or software repositories to bridge technical and non-technical perspectives (16–19). Tensions emerge around shared understanding, particularly when new research questions surface that 'break' initial common ground, and when data are obscure or inaccessible, they become difficult both to make sense of and to communicate to others (20). Meanwhile, research on healthcare collaboration has examined partnerships between researchers and decision-makers, revealing benefits such as mutual understanding of each other's realities, alongside costs including additional time demands and trade-offs between academic recognition and practical problem-solving (21). Recent work on AI in healthcare has developed frameworks for implementation-stage collaboration (22–26), with organization and structure suggestions for team building including multiple disciplines, as well as high-level stages of working. Researchers have also examined power dynamics between multiple stakeholders when AI systems are enacted in care settings (10,27,28), a step much further ahead than the development.

However, both literatures leave a critical gap: the formative stages of AI development in healthcare research remain underexplored. While frameworks provide useful ways of thinking about AI collaboration in healthcare, they do not reveal how the work is actually achieved in practice (23). Fitzpatrick and Ellingsen (29) also found that the overwhelming majority of computer-supported cooperative work (CSCW) research in healthcare focused on workplace settings. While calling for more research in formative stages, they explicitly mentioned only policy-making and procurement rather than actual development work. This gap matters because the challenges identified in data science likely manifest differently in healthcare research, where clinical expertise, patient involvement, and regulatory considerations add layers of complexity. AI development specifically amplifies these challenges in two ways. AI's computational complexity compounds the problem of obscure data. Teams must navigate data access issues, and the algorithms themselves function as 'black boxes' (30) that are difficult to explain and communicate across disciplinary boundaries. AI's requirement for large datasets creates practical challenges around data access and building shared understanding of what these datasets contain, particularly in healthcare where data are often sensitive and require clinical expertise to interpret. Examining the actual work practices during early AI development reveals how these AI-specific challenges manifest in healthcare research teams.

To fill this gap, we explore the collaborative AI practices of three research consortia through a series of semi-structured interviews. Each consortium is tasked with developing and applying AI algorithms to improve the management of multiple long-term conditions (multimorbidity) in the UK (31). We understand *collaborative AI practices* to be the work that takes place around data and the development of AI algorithms. A consortium is typically understood as a group of organisations joining together for a shared purpose (32), and the participants in this study consist of academics, public health experts and patient representatives. The consortia are funded by the National Institute for Health and Care Research (NIHR), which explicitly states that all of the research must take a 'multidisciplinary' approach (33).

We argue that the processes associated with the combination of AI and healthcare research introduce new obstacles to interdisciplinary collaborations. We therefore seek to address the question: how do AI work practices complicate ways of interdisciplinary work in healthcare research?

By exploring the interactions between the stakeholders during the development of AI algorithms, we contribute with the following:

- We expand the human-computer interaction and CSCW literature by focusing on the formative stages of AI algorithm development in healthcare and uncover how the addition of AI impacts the confluence of healthcare and interdisciplinarity research

- We break the convention of pitching data scientists against single-domain experts by including perspectives from clinicians, statisticians, and staff involved in patient and public involvement and engagement (PPIE). In doing so, we examine how consortium participants described the communication and coordination practices involved in bringing clinical and patient perspectives into AI development.
- Overall, we show how the development of AI algorithms, whilst being a novel and exciting technology, can add new difficulties to the already known problems of interdisciplinary teamwork.

# Study design and methodology

## Consortia at a glance

All three consortia are funded under the Artificial Intelligence for Multiple Long-Term Conditions NIHR call (31). Each consortium employs between 20-30 academic and healthcare researchers. Some, generally early career researchers, were employed full time, and others, typically more senior academics, were employed part time on the projects. The exact number of people employed by each consortium fluctuated over time as people joined and left the project. Additionally, each consortium engaged with members of the public and patients (PPIE) who provide lived experience expertise to the research projects and were reimbursed an hourly rate for their contributions. A typical consortium spans various work packages (WP), with a combination of researchers from different disciplines in each WP. For example, a typical WP includes epidemiologists, data scientists, clinicians, social scientists and PPIE contributors. Cutting across all WPs are administrative staff, including PPIE coordinators, who typically organise and manage all PPIE activities. It is common that, across a consortium, researchers are based in different geographical areas, at different universities and that work takes place both online and in-person. Common themes across the consortia are the use of large datasets, particularly electronic healthcare records (EHR) collected from primary and secondary care in the UK and the application and development of AI methods in research (31). Although there is no agreed on definition for AI, the NHS's states that 'AI is the use of digital technology to create systems capable of performing tasks commonly thought to require human intelligence' (34). In the consortia in this study, AI refers to complex mathematical models that can be used to analyse data, for example, to make predictions about future health problems, or identify groups of similar patients. For example, one consortium uses EHRs to identify clusters of patients that share similar health trajectories, and then predict patients movements between different clusters as diseases progress (35). Whilst some consortia have long-term goals of integrating their findings into clinical practice, most of the work in this stage of funding is on the development of AI algorithms and discovery of new knowledge (e.g. associations between diseases and drugs) rather than the implementation of AI algorithms in care settings. In our study, we focus on the interactions among stakeholders as they process data, decide, build AI algorithms and validate and share findings. This setting is important for interpreting the findings as the consortia were UK-based, used health data generated within or linked to NHS care, and were operating within UK research governance, data access, and PPIE expectations.

## Participants

We recruited 13 participants across 3 different consortia, identified as A, B and C, between January 19th 2023 and March 3rd 2023. A mixture of convenience and purposive sampling was used. Our sample size of 13 was arrived at using Malterud et al.'s (36) notion of information power, which assesses qualitatively how much information a single sample (or participant) contributes with across five axes. First, the aim of the study was narrow focusing specifically on collaboration in interdisciplinary AI research teams. Second, the sample specificity was dense as each of the participants had considerable knowledge of the area of interest. Third, this study is based on established works from CSCW, HCI, and the social sciences more broadly. Fourth, the quality of dialogue is strong, with clear communication between researchers and participants. Fifth, analysis is cross-case with participants approached with diverse experiences due to their disciplinary backgrounds. The recruitment of participants started with an email sent to all the internal, employed, participants across the consortia inviting people to take part. Any member of the consortia that were reachable by email or indicated by other members was eligible. Then, to ensure a range of disciplines were represented, individual emails were sent to recruit individuals with specific roles (e.g. someone with a clinical background). There were a higher number of individuals from a data science related discipline (data science, data engineering, statistics) recruited, as these made up the majority of the employed staff in the consortia. Across all the participants recruited, five worked full-time on their consortium, while the rest were part time. Six of the interviewees held a leadership position either in a small group or a work package. Our sample included PPIE coordinators but did not include patient or public contributors themselves; findings concerning patient perspectives therefore reflect how these perspectives were described by consortium members and PPIE coordinators. Table 1 summarises participant information.

Table 1: Description of participants.

| **Name** | **Role** | **Consortium** | **Access to data** | **Primary data analysis tools** | **Leadership role** | **Full or part time** |
|---|---|---|---|---|---|---|
| P1 | Academic GP | A | Direct | | No | Part |
| P2 | Data scientist | B | Direct | | No | Part |
| P3 | Epidemiologist | A | Upon request | | Yes | Part |
| P4 | Statistician | A | Not needed | R, R Markdown | Yes | Part |
| P5 | Data scientist/AI Coordinator | B | Direct | R, Python | Yes | Full |
| P6 | Data scientist | B | Direct | | No | Full |
| P7 | Postdoctoral research associate (Data science) | B | Direct | Python | No | Full |
| P8 | Postdoctoral research associate (Data science) | B | Direct | Python, Jupyter notebook | No | Full |
| P9 | Data scientist | B | Upon request | MATLAB; Python | Yes | Part |
| P10 | Clinician | A | Indirect | R | No | Part |
| P11 | Academic clinical fellow | C | Direct | STATA | No | Part |
| P12 | PPIE coordinator | C | Not needed | | Yes | Part |
| P13 | PPIE coordinator | B | Not needed | | Yes | Part |

## Interviews

An initial interview guide was developed by RH and ER based on existing literature across CSCW and HCI (16–20) and our own experiences of interdisciplinary collaborative work and later refined through input by the wider research team and a pilot interview.

Interviews were conducted by RH and ER. At the time of the project, RH was a postdoctoral researcher with experience in data science and ER was a PhD candidate with experience in epidemiology and PPIE. The interview guide encouraged participants to reflect on 5 main areas; (1) the participants role and team structure in the consortium, (2) data access or interaction with data, (3) data analysis tools and work practices with data, (4) methods and approaches used personally and within their team for sharing knowledge and (5) experiences of communicating knowledge to other members during the research. We verbally collected basic descriptive information about each participant including role, full or part time employment, consortium name, level of access to any datasets and specific data analysis tools used. Demographic details were not collected as these were not considered relevant to the study aims and given the small sample size reduced the risk of re-identification of study participants.

The 13 semi-structured interviews were scheduled for 1 hour each. This consisted of 9 face-to-face interviews, and 4 remote interviews conducted via Microsoft Teams. All interviews were recorded and immediately transcribed into text using otter.ai (37). The transcripts were further corrected manually when the tool did not identify the correct words.

Ethical approval for this study was granted by Queen Mary Ethics of Research Committee (ref. QMERC22.341). All participants were supplied with an information sheet and signed (digitally or physically) written consent forms. Participants were reimbursed with a £25 voucher for participation.

Participants were invited for study interviews only and were not involved in the coding or discussion of results.

## Creation of codebook and analysis

Themes were analysed following Braun and Clarke's guidance for thematic analysis (38). Both RH and ER attended all the interviews so were familiarised with the content. Transcripts were coded in an iterative manner utilising Taguette (39). Independently both RH and ER coded four transcripts, selected to represent a range of participants from different disciplines. Over multiple sessions we agreed on an initial codebook, harmonising the vocabulary and refining codes. Both RH and ER then jointly revisited the initial four transcripts in order to revise codes and add additional codes as needed. The remaining transcripts were then divided, RH coded 5 and ER coded 4 transcripts independently, meeting to discuss modifications and additions to the code book. This approach aligns with Braun and Clarke's reflexive thematic analysis, which is grounded in an interpretative paradigm and does not advocate the use of inter-coder reliability statistics as maker of rigour

# Findings

In this section, we present the three themes that we identified in the analysis, summarised in Table 2. We discuss how participation in AI research is often novel for participants resulting in different expectations of the capabilities of AI and how this novelty can also amplify the differences between disciplines as stakeholders learn new ways of working. We identify how and where stakeholders communicate about AI and how they ensure their messages get across. Finally, we discuss the benefits and challenges of combining multiple real-world healthcare perspectives in the form of clinicians and PPIE coordinators in these highly diverse teams. We highlight in **bold** the subthemes that we identified in the analysis.

Table 2: Table of themes and subthemes

| **Theme 1: AI novelty** | |
|---|---|
| *Navigating unknowns*<br>How stakeholders navigate the unknowns of AI when they are not familiar with it | 'But yeah, we'll have to see how it works. Because I think my impression is not new, just for me, I think it's, it's new to everyone who's working in, in the work package.' (P10)<br><br>'Unfortunately, my machine learning background doesn't allow me to always understand everything. But as I said, everyone is actually quite good at trying to explain things in a way that I can make sense, even though I don't understand mathematically how you get from A to B, I |
| | can still understand the thought process behind it.' (P10) |
| *Disciplinary differences*<br>Stakeholders face different workflows and expectations compared to what they were previously used to in their disciplines | 'I generally don't have problems with them presenting whatever they think is the most relevant aspect there, which may be 15 different slightly ways of clustering that they have done because they're trying to experiment something in particular, I tend to cringe way more if they attempt to do the same on the monthly meeting where the clinicians are there because what, yeah, but what's the point? I mean, where is the conclusion? For us computer scientists, the conclusion maybe I've tried a lot of very interesting things. But for clinician the point is, which one of these would you recommend for use?' (P5)<br><br>'Epidemiologists have certain perspectives about modelling and you try to make sense of them so that they are meaningful, interpretable and practical. I say practical, it's easy to come up with all sorts of models, but if it results the same as a simple model, why bother?' (P3, Epidemiologist) |
| *Computational complexity*<br>Challenges that the complexity of AI introduces to the interactions between stakeholders | 'So some of the groups can work in different ways and some are within National Safe Haven and they're limited in what they can do. And it also affects the AI because obviously those National Safe Havens, we are now fighting to get a little bit more into the environment. But someone has to install it for you. Someone has to approve it. And they don't have GPUs which means we are seriously stuck in terms of deep learning. The Safe Haven currently procured two GPUs for all the projects. I mean, we would use 4 just for ourselves, that's two shared with another 50 projects.' (P5) |
| **Theme 2: AI knowledge sharing** | |
| How stakeholders adapted (or not) their communication depending on their audience's knowledge of AI | 'if I'm discussing with programmers, [...] I keep it [the notebook] live because people may ask, can you change this and show me.' (P5, data scientist).<br><br>'[With clinicians] I'd usually share it kind of, like, processed data by Excel or shared |
| | sheet, to kind of minimise the time they've got to spend on it if they're doing something for my benefit.' (P11, clinical fellow)<br><br>'For the kind of wider group presentations, I always say, give a presentation on slides' (P12, leadership) |
| **Theme 3: Multiple real-world healthcare stakeholders** | |
| *Sense checking*<br>Importance of clinical and lived experience to validate findings and guide AI research | 'the clinicians involved in the project, they verify that what I'm doing makes sense and there's an actual utility in practice.' (P8).<br><br>'And the useful thing, especially when talking with clinicians, is that they provide a lot of practical examples, so like, we might present an abstract model and they come up with a specific example from the real life, they say, you know, maybe this is useless because this never happens in real life or maybe you have to take into account the special case because it is very common in practice.' (P8) |
| *Relevance to patients*<br>Importance of patients' views to ensure long-term utility of AI findings | 'So we've had some really nice interesting things like that, that I'm not sure would have come out if it was just the group around the table of data scientists They will be, you know, more talking about the differences between the different models, rather than the kind of step back for what does this mean about modelling for patients.' (P12) |

# AI novelty

Although the term 'AI' has been around since the 1950s (40), the popularity of AI has grown exponentially. For many stakeholders these consortia provided their first exposure to developing AI algorithms, requiring them to **navigate the unknowns** of AI.

> 'But yeah, we'll have to see how it works. Because I think my impression is not new, just for me, I think it's, it's new to everyone who's working in, in the work package.' (P10, clinician)

Given the novelty of AI for many stakeholders this resulted in a certain confusion and a lack of shared understanding of AI methods. Often AI methods were construed as complex 'black box' methods that even experts, or stakeholders working hands-on creating the AI algorithms, failed to understand. 'Black box' is a label typically given to deep learning methods, a branch of AI methods whose internal mathematical processes are not easily understandable and therefore it is difficult to interrogate how the AI method produces the output. These AI methods were hard to understand for stakeholders in disciplines outside of data science (such as clinicians), due to collaborators' different disciplines, knowledge or experience with AI.

> 'I'm not like a statistician, you know, so I wouldn't necessarily know all the different, like, you know, AI methodology so, I don't know, but it depends what the conversation is. So if someone was specifically to have a conversation with me to need my input on something, and then I'd probably want to know a bit more about what exactly they were doing' (P1, clinician)

Stemming from the lack of experience, collaborators often held different expectations of what AI algorithms were capable of:

> 'I came to this PhD, you know, reasonably naive about what AI is, and I've had to kind of develop my explanation to what it is for, you know, for other healthcare professionals and for the PPIE group as well. So it's definitely a challenge, and with all of the kind of various ideas of what AI really is, I think it leads to kind of different expectations of what it can deliver. So kind of everybody thinks, 'oh, wouldn't it be amazing if they could do this, this and this', when the reality of it you know, either what we're working on or what is even possible is not there.' (P11, clinician)

These expectations of AI impacts collaborative AI practices, as they need to be managed for the project to move forward. Expectations also impacted what data scientists were asked to do, which at times differed from what was computationally possible. Stakeholders needed to develop a level of shared understanding of the capabilities of AI which was typically achieved by communicating the abilities and limits of AI models:

> ‘It is good to know, let the clinicians know that all the models don’t work every time so they a little bit of those, giving them some idea that yeah, what models are and what my model will do and what is my models limitation.’ *(*P2, data scientist)

The field of computer science and development of AI algorithms is one of rapid change that appears distinct from other statistical or big data disciplines. New methods and models are constantly being released, and there is a preference to publish in conferences rather than academic journals, as this is typically a quicker way to publish (41). This culminates in a field that is fast-paced and values novelty over practical implementation, which is often in direct contrast with the work practices of other stakeholders, leading to **disciplinary differences**:

> ‘I generally don’t have problems with them presenting whatever they think is the most relevant aspect there, which may be 15 different slightly ways of clustering that they have done because they’re trying to experiment something in particular, I tend to cringe way more if they attempt to do the same on the monthly meeting where the clinicians are there because what, yeah, but what’s the point? I mean, where is the conclusion? For us computer scientists, the conclusion maybe I’ve tried a lot of very interesting things. But for clinician the point is, which one of these would you recommend for use?’ (P5, data scientist/AI coordinator)

Due to the rapidly evolving field of AI, it is common to try different models and approaches to the analysis to find the most appropriate or best-performing model, as highlighted in the previous quote running 15 variations of the same analysis. This can be in contrast with other disciplinary practices:

> ‘So the emulation of a trial would involve specifying a specific protocol that you’d follow quite closely. So there should only be one set of results. It’s not like we’re going to be playing around with the results […] we’re not going to be discussing about changing the analysis after we’ve done it because the point of it’s meant to be kind of pre specified’ (P4, statistician)

These differences in disciplinary experiences also fed into collaborators having different approaches to statistical modelling:

> ‘Epidemiologists have certain perspectives about modelling and you try to make sense of them so that they are meaningful, interpretable and practical. I say practical, it's easy to come up with all sorts of models, but if it results the same as a simple model, why bother?’ (P3, epidemiologist)

For some collaborators, AI offered endless possibilities and solutions to any problem, whilst for others, AI needed to outperform more traditional and simpler methods to demonstrate value. For stakeholders that had previously worked with developing AI algorithms there was also a frustration around the limits of AI methods especially around concepts that were important or general practice in other disciplines:

> ‘The other thing is a confidence interval like an uncertainty measurement, it doesn't exist among AI people. So there's no method and I think my colleagues have been doing something about it, it's new work. I don't know whether, to what extent that will be applied, but we have to address that requirement of clinical medical health researchers because that's really intrinsic. You can't have no measure of uncertainty.’ (P3, epidemiologist)

Developing AI algorithms not only added methodological complexity, but also **computational complexity** to these projects. AI methods, and in particular deep learning, require huge computational resources to run, typically using multiple graphics processing units (GPUs). GPUs are specialised devices that allow for very rapid mathematical calculations and are costly, both financially and environmentally.

> ‘So some of the groups can work in different ways and some are within National Safe Haven and they’re limited in what they can do. And it also affects the AI because obviously those National Safe Havens, we are now fighting to get a little bit more into the environment. But someone has to install it for you. Someone has to approve it. And they don’t have GPUs which means we are seriously stuck in terms of deep learning. The Safe Haven currently procured two GPUs for all the projects. I mean, we would use 4 just for ourselves, that’s two shared with another 50 projects.’ (P5, data scientist)

This requirement and limitations for computational resources sits alongside the need for data safety. EHR data includes sensitive patient data and at times it is provided through Trusted Research Environments (TREs) or Data Safe Havens. Safe Havens are machines that can be accessed remotely and often all modelling and analysis must take place within this environment so that data files cannot be taken out of it. This predetermined environment can limit the number of GPUs available, but also the types of programming languages available, which restricts what work can be done:

> ‘There is only R in that particular National Safe Haven’ (P5, data scientist).

The development of AI algorithms in healthcare research adds complexity to collaborative AI practices. AI is often a new topic for collaborators, particularly those outside of computer or data science, and people need to develop new working styles to accommodate the uncertainty of AI methods. Stakeholders from different disciplines may have expectations of what AI methods can do which are not always fulfilled, and certain disciplinary demands mean that new forms of AI need to be created.

## AI knowledge sharing

Given the requirements of working on AI algorithms in healthcare, including data preparation for modelling, interpretation from clinicians and project demands such as progress updates, we wanted to understand how participants addressed or considered the diversity of disciplines and backgrounds within the consortia in their collaborative data practices, such as preparing materials for sharing findings or obtaining feedback. We asked them to think about their practices when sharing information about AI as well as their perception of how AI was communicated by others in the project.

Participants reflected on the settings for collaboration, such as the places where it happened or the media used to facilitate it. The complexity of AI and the structure of projects, including the multiple disciplines and their organisation in work packages, led to a variety of tools being used to collaborate and share knowledge, from emails to online repositories such as GitHub (a platform for managing and sharing code).

We identified that participants related the novelty and complexity of AI to the **background** and **perceived knowledge** of team members when they were choosing the tools for sharing and content of shared materials. Within this there were stark disciplinary differences in the typical tools used for sharing of information. Data scientists often used GitHub; this was, however, seen as introducing obstacles to other disciplines:

> 'where some people have shared GitHub repositories, my personal view is that [...] data scientists [...] might find it very easy to navigate, but in reality for a patient, it's just a whole other thing that [...] I don't see that as a huge benefit.' (P12, PPIE coordinator)

Code sharing aside, participants described meetings as the main setting for collaboration in giving progress updates, asking questions and obtaining feedback. They paid close attention to full team meetings, which included 20-30 people from all disciplines and which were seen as an opportunity to interact with people who were not engaged full-time in their projects, such as clinicians. In some consortia, participants also detailed smaller meetings, for example, where all data scientists would join, or with specific colleagues when problems could not be solved in the larger meetings.

Most data scientists described performing their work using R or Python as programming languages. For this, they used popular software such as Jupyter Notebooks that combine text editing with visual outputs. Participants described using these tools and sharing the notebook documents with people who had the same background or role as them:

> ‘if I’m discussing with programmers, [...] I keep it [the notebook] live because people may ask, can you change this and show me.’ (P5, data scientist).

The interactivity that the Jupyter Notebook platform allows was suitable for this kind of collaboration, but participants were careful when using it to not overwhelm their peers from other disciplines and they were also aware of the limited time for updates in meetings. When they communicated with members from different disciplines, data scientists often moved data content from programming-oriented tools to text documents or spreadsheets. The use of accessible tools was sometimes imposed by Data Safe Havens, from which data and results cannot be transferred out without a request to the administrators.

> ‘[With clinicians] I’d usually share it kind of, like, processed data by Excel or shared sheet, to kind of minimise the time they’ve got to spend on it if they’re doing something for my benefit.’ (P11, clinical fellow)

When deciding what to share in meetings, participants considered the background of audience and the format of the meeting, and then adapted materials accordingly:

> ‘Each of these meetings, they have different audiences. And it's not the same, like a meeting that I do with these people working on AI and a meeting with everyone in the project, with all the clinicians […] then I have to quite tailor the way to communicate and the things that I'm showing that are important. I don't go that much into the maybe some methodological hard details, but I keep ideas […] (at a) higher level’ (P7, data scientist)

Here, participants describe the challenge of deciding how much detail to convey from the AI work, from the mathematical details of models to a very high level overview of their work, and, once the content is chosen, how to communicate it. There was also an awareness, that there are perceived limits to what people could do to facilitate communication across disciplines:

> ‘I think there’s a real variation amongst individuals and amongst their kind of home discipline … as to what point, they sort of say, well, you know, “it can’t be any more simplified than that … I’ll just say this particular graph theory” … and it’s at that point that I might then go away and say, and try and find a simple description or something on the internet … that hopefully helps to bridge the gap’ (P12, PPIE coordinator).

Beyond attending to the disciplinary affiliation of their colleagues, participants used this information to inform their perceptions about how much knowledge colleagues would have about AI.

> ‘If we’re doing a presentation to a PPIE group I might keep what we show quite limited and have more text to explain what the results show in, in an accessible way as possible. If I’m talking to you know, with statisticians, we might have more plots’ (P4, statistician).

Hyperparameters are settings of computational methods that can be manually defined by data scientists or through automating tools. The data scientist believed that these details about the method were unnecessary for clinicians, whereas the statistician thought certain plots were more appropriate for other statisticians but not accessible enough for patients. However, not all participants felt the needed to adapt materials when collaborating:

> ‘I’d say that for the clinicians collaborating with us. They are so experienced with machine learning and artificial intelligence […] So we don’t really need to use lay language to be honest with you. So yeah, they are they are familiar’ (P6, data scientist)

In this case, the participant knew that those clinicians were familiar with AI methods, therefore they did not see the need to adapt their communication method. However, in large meetings there were people from different work packages and participants did not always know who was experienced with AI and who was not. Participants also had limited time to discuss during meetings, which also influenced what they wanted to share:

> ‘I think it’s just the balance of how much explanation you have and how long, I mean, if you’re in a full team meeting, you probably don’t have as long, [as] if you’re in a work package meeting’ (P4, statistician)

Participants described a complex picture of AI knowledge sharing, with various decisions taken in light of the limitations and the difficulties of doing AI work in healthcare. Common data science workflows were disrupted by the requirements of healthcare data, whilst the key collaboration settings, such as meetings, restricted the amount of time for collaboration and required filtering and adapting content for multiple disciplines.

## Multiple real-world healthcare perspectives

All the consortia in this project relied on large EHR datasets. EHR datasets are collected in healthcare settings often as part of routine care. Patients participate by generating data every time they interact with the healthcare service. These interactions include diagnoses made by clinicians or blood tests taken by other healthcare professionals, which are then stored in EHRs. This data is later analysed by data scientists and offers a representation of patients' experiences of health. By real-world healthcare perspectives, we refer to clinical experience and to patient perspectives as they were described by consortium members and PPIE coordinators. Contrary to the researchers who only see the collected data, clinicians and patients have first-hand experience in the creation of the data as well as taking part in its analysis.

Although it is common in healthcare research to include clinicians, it is more unusual to have patient stakeholders involved, and both stakeholders help to add real-world healthcare context and highlight the subjectivity of the data. Subjectivity of big data has been well researched and underscores that data is never ‘raw’ (42) and this approach highlights the socio-technical contexts that shape and create data (43). EHR data is inextricable from the setting where it comes into being, and patients and clinicians in interdisciplinary settings can help spotlight this whilst aiding in **sense checking**:

> ‘Most of the people involved in the data analysis have no background on mental health, so as much as they can do a quick Google search, and say, oh, I wonder if this is why this is like this, they don’t have the clinical background. That’s why we got involved in this project, to actually provide that. And some things that might be very obvious to us might not be very obvious to data analysts. I mean, the data is there, but there might be some relationships that we’re like, oh, yeah, that makes sense from the clinical point of view’ (P10, clinician)

This sense checking relies on the practical experiences of clinicians, sometimes both the clinical knowledge, but also the knowledge of how the data was collected and what it might represent. Although data scientists have access to the data directly, and can look for data driven insights, it still relies on checking with others who have clinical or lived experiential knowledge.

> 'And the useful thing, especially when talking with clinicians, is that they provide a lot of practical examples, so like, we might present an abstract model and they come up with a specific example from the real life, they say, you know, maybe this is useless because this never happens in real life or maybe you have to take into account the special case because it is very common in practice.' (P8, data scientist)

PPIE coordinators also played an active role to ensure that these projects had **relevance for patients:**

> 'Like I said, the challenge is that patients quite rightly jump to: what does it mean in practice? And we know there's a gulf between an initial data analysis where we think we've got an indication of the prevalence of whatever, to what that will ultimately look like in the tool.' (P13, PPIE Coordinator)

The involvement of patients in the development of AI algorithms is not standard practice, although it is mandated by the NIHR for all their funded projects. Patients could indirectly influence data scientists working on the projects, through communication in feedback sessions or presentations. PPIE coordinators emphasised the impact of the AI research on healthcare practice or the impact on patients as individuals in the healthcare system. PPIE coordinators reflected on the role patients played in the consortia by reshaping collaborators' perceptions of the data, moving from viewing EHR datasets as anonymised, decontextualised datasets, to seeing the data as 'personal Big Data' (44), with each data point and graph representing a patient and their health experiences. Although the data used in these projects may or may not include data from the patients that are part of the PPIE groups, these patients are still engaged in making sense of this ‘personal’ data as if it were their own and ensuring the best outcomes for patients.

> ‘So we’ve had some really nice interesting things like that, that I’m not sure would have come out if it was just the group around the table of data scientists They will be, you know, more talking about the differences between the different models, rather than the kind of step back for what does this mean about modelling for patients.’ (P12, PPIE coordinator)

However, sometimes the preferred focus of the patients and what they would like to make sense of was not always able to be met by data scientists. Commonly there was little or no data available in the datasets used by these consortia for the concepts that patients were most interested in, such as quality of life, or activities of daily living.

> ‘We do have discussions with our PPIE members about what are the important findings for them, and that is important or interesting, because sometimes, for example, we had a call where they very much spoke about quality of life. And that's not something that was directly measurable in the data, but that's how they saw the next stage of the research going, saying, can we find this? But how does that actually impact me? And so I think the researcher was kind of left like, okay, I need to go away and think a bit more about what does this actually mean going forward? So how can I look at some of these things that people have brought up? That may not be directly in the data, but are there kind of proxy measures and things that we can use?’ (P13)

The process of capturing disciplinary expertise to shape AI development is not easy and involves working at the intersection between disciplines. These consortia include multiple real-world healthcare perspectives, clinicians and patients, who are familiar with the context of where the data is created that goes onto feed AI algorithms. This allows for not only sense checking between collaborators but can also act as a reminder of where the data are from, and that they are a representation of the real patient health experiences. This work helped to personalise big data and, at times, shape the direction of the work to maximise impact for patients.

# Discussion

In this section, we reflect on the key findings concerning our primary motivation for this work and how these findings compare to the literature. We also discuss the limitations of our approach and future research opportunities.

## Obstacles to collaboration

In this work, we identified that the development of AI algorithms in interdisciplinary research exacerbates many existing tensions and challenges in collaborations. Taken together, our findings support the sociotechnical framing that guided the study (11). AI development was shaped not only by technical factors such as computational resources, data access, and coding environments, but also by social factors such as disciplinary norms, working practices, expectations, and collaborative routines. This was especially visible in the way AI amplified existing challenges around common ground, communication, and decision-making in interdisciplinary healthcare research.

Many themes highlighted in this work are mirrored in large scale academic research projects or healthcare collaborations, what this study adds is how the development of AI algorithms impacts this highly interdisciplinary work. As always there are differences or tensions in the common ground between disciplines as discussed by Mao et al. (20). Common ground has been extensively studied and refers to the overlap in knowledge collaborators have about others and the task. In this work, we found that disciplinary differences led to a lack of shared knowledge which was particularly heightened by the novelty of AI. For many researchers this was the first experience of working with a team to develop AI algorithms for healthcare problems. Due to this, individuals had differing expectations and understandings of AI, and at times wanted to take approaches that were unfeasible or impossible. Data scientists were often aware of differing expectations from other disciplinary team members, and this shaped how they talked about their AI methods in order to set realistic expectations. Prior work has also revealed the need to set expectations when convincing stakeholders about the importance of AI (16). In the consortia, often this meant that data scientists presented numerous alternative models or results in meetings, which possibly helped set expectations but hindered quick decision making and the progression of the consortium's work.

Another common problem the consortia faced was the challenges around access to large electronic healthcare records or health data. Health records are personal data and often subject to rigorous ethics committee approvals, whether data has been anonymised or not. Depending on the data provider, health data may also only be available for research via Data Safe Havens or TREs. In these settings, there can be restrictions on the work, due to computational or technical restraints and outputs often need to be verified before being transferred out to public spaces (i.e., a researcher’s computer). These aspects make it more difficult to prepare and use large scale health records in research (45). As AI methods rely on ever-growing datasets, this has amplified many of these challenges, having a knock-on effect on how collaborative work is done. The infrastructure needed to store large datasets, such as Data Safe Havens, influences how people could work with the data, reducing access to a handful of approved people. Individuals needed to engage in additional work to prepare information to share with those unable to access the data, in ways that meet the requirements set by the data provider, for example by making slides, summaries of the data, or fictional case studies that could be discussed. This often slowed down collaborative work, as additional steps were required, at times even fake or ‘dummy’ records needed to be made so that work could be discussed with those without approval to access the data.

The computational complexity of AI methods also brought challenges to the collaborative processes. Similar to other collaborative work, meetings were a prominent setting where multiple disciplines met to discuss work and provide feedback. However, as AI methods, particularly those using large datasets, often require significant time to run, participants reported not being able to give prompt updates after receiving feedback. This led to slower development, as data scientists needed to go away, run new models and calculations and then return to another meeting with new results. Setting expectations of AI, as described above, was a way of mitigating this to avoid wasted time.

## AI knowledge sharing

Our results suggest that the development of AI algorithms impacted multiple collaborative processes, from how individuals accessed data, to giving feedback and communicating data and results.

For those communicating data or AI methods and results, this impacted both how and what was communicated. We identified a transition from programming-oriented tools to general software such as Excel and PowerPoint. This was a key point where data scientists diverged from their typical disciplinary work in order to communicate with other team members. This has parallels to prior work which highlights how data scientists engaged with management in companies and clients (16,19,46,47). In contrast to previous work, where AI communication was centred around convincing collaborators and pitching ideas, in the consortia, the communication around AI was towards the common goal of progressing the consortium's daily work, and focused on presenting current challenges with data or results from AI algorithms. Data scientists needed to adjust their communication styles to the different disciplinary backgrounds of others and to different collaborative settings such as meetings. New communication tools have been suggested (16) to facilitate better and easier communication between disciplines around data, but it is likely that tools alone are not enough In the context of academic research, movements towards transdisciplinarity or team science (6,48) could bring benefits to promote closer interactions between disciplines. It must also be noted that, as data scientists had to learn to communicate complex ideas to different audiences, these different audiences also had to learn new knowledge, methods and languages used by other disciplines. Communication adaptation is therefore a collaborative process of learning by all involved.

It is not yet clear how much the communication process itself suffers from this disruption, and the constant need to convert information into a shape suitable to share between disciplines. As data scientists need to decide which content to share based on their own expertise and their perceived knowledge of others, it is possible that important information is lost in this step between analysis and sharing (17). As communication with stakeholders flows in both directions, it is also reasonable to assume that information is also lost on the way back. Research that covers the whole AI development could determine the extent to which information is lost and how to mitigate it, for example, through new tools that address disruptions or a different kind of project management that enables data scientists to preserve their workflows and for others to have oversight of it at the same time.

## Additional perspectives in collaborative AI research

The consortia in this study presented an opportunity to explore how multiple academic disciplines, clinicians, and PPIE coordinators described the incorporation of clinical and patient perspectives into AI development. Because patient and public contributors were not directly interviewed, our findings should not be read as direct evidence of their experiences, but rather as evidence of how consortium members sought to include, translate, and respond to these perspectives. Previous research in this area has typically only looked at data scientists and clinicians (or medical experts) and has not explored how other stakeholders, such as patients, can influence AI work practices. Through the views of the PPIE coordinators, we found that both PPIE participants and clinicians exerted a significant influence on collaborative work practices, from what and how materials were shared, to the language used and how others perceived and interacted with the data.

With the increase of the development of AI algorithms for healthcare, there has been a rise in the inclusion of patients and members of the public throughout the development lifecycle (49–53). Previous work in AI, although not explicitly in healthcare research, has highlighted the benefit of working with patients to reduce biases (49,50), and to improve trustworthiness of AI algorithms (54). We found that AI-in-healthcare research can add additional layers to PPIE work, requiring new roles and work. In these consortia, PPIE coordinators often acted as 'translators', engaging in additional work to help researchers effectively communicate their research to patients, whilst trying to minimise the complexity introduced by AI and the language around it. This active role has been described before as knowledge brokering (55,56) but is often assigned to clinicians in healthcare settings. In the consortia, PPIE coordinators often had a better understanding of what information and format was appropriate to share with PPIE participants and were crucial in clearing the muddied waters around AI, thus ensuring that the real-world perspectives were actually being taken onboard by the consortia. This mirrors prior work that highlighted the influence of power dynamics and the importance of facilitating knowledge transfer between disciplines during collaborations (27,57). We found that researchers were relatively unaware of the volume of additional work by PPIE coordinators to allow materials to reach a stage where they were suitable for patients.

Patients and clinicians also added additional perspectives to the collaborative teams, which are not commonly present in collaborative AI research. This study highlighted that they aided in a grounding of the data, helping to relate it back to the settings under which it was collected and remind other stakeholders of data's inherent subjectivity. These voices also acted as a guide for AI research, driving the project towards what is relevant for clinical practice and patient experiences. As academic funders and private companies are increasingly interested in involving patients in research (58,59); balancing the requirements of genuine participation in AI research and the nature of academic interdisciplinary collaborations remains a challenge. For HCI and CSCW researchers, these developments open up new opportunities towards influencing the development of healthcare, as called by Fitzpatrick and Ellingsen (29), particularly to empower these knowledge brokers, and thus, patients and the public.

# Limitations and future work

The study presented here is a first step in understanding how AI impacts collaborative work practices between multiple disciplines. We acknowledge that there are various limitations to this study and offer opportunities for future research.

We recognize the unique context related to the consortia in this study. First of all, the datasets used in most projects come from the National Health Service (NHS) in the UK which presents its own challenges for data access and restriction. Our findings may not be relevant to work using other datasets or citizen science projects (60). Secondly, as the consortia all operate within the UK they are therefore embedded in local research practices and culture who could influence work practices. For example, in the UK, hierarchy is entrenched in modern healthcare systems (61) and academic research groups (62), which is likely mirrored in these consortia and may impact collaborative work.

All of the consortia in this study were still in their first year of full implementation at the time of the interviews, and many only just had access to the data. At later stages of the projects, we expect that some of the themes identified in this research may be more or less relevant therefore a follow-up study would be of great value to complement this work. However, feedback from participants at earlier stages of research projects is rarely reported in the literature and allowed us to explore some of these themes whilst they were still fresh in participants minds. Our work also opens paths to in-situ scenarios for research on adapting and developing tools centred on AI communication. These scenarios include balancing the information needs of stakeholders against the limited amount of time that more senior academic researchers tend to dedicate to each project.

We did not interview any patients that were part of the PPIE groups, although we did interview PPIE coordinators. This was partly due to time restraints of this study, as many consortia had not yet started working closely with their PPIE groups and the time needed for us to develop and test interview materials to ensure their suitability for patients. Nevertheless, we did manage to expand previous work that typically focused on data scientists and clinicians only by including multiple other perspectives on the data science workflow and interdisciplinary work.

# Conclusion

In this paper, we presented the results of a thematic analysis of 13 interviews with members of 3 large research consortia working on the development of AI algorithms in healthcare. Our findings highlighted how AI introduces new obstacles to collaborative AI practices, from data analysis to sharing of results. We revealed the key impacts of AI in how participants approached their consortia, in particular the newness of AI and the increased differences between disciplines. We also learned how the complexity of AI as a topic influences the creation and sharing of artefacts to progress work.

Our paper joins the growing body of literature on the collaborative practices of interdisciplinary teams focused on data science research. Although additional work is required to develop detailed recommendations to support interdisciplinary collaborations in AI and health research, we offer preliminary recommendations for academics and practitioners interested in AI-in-healthcare projects. Firstly, setting early and appropriate expectations about AI capabilities and its limits is crucial, this could be achieved through developing AI literacy for all stakeholders. Considering the role of PPIE coordinators and the participation of public and patients, support for translating technical outputs into accessible formats is also essential. This may be in terms of training for stakeholders in how to communicate their work to lay members, as well as assistance and time converting documentation into easy-to-read formats. Finally, funders and project leaders should also recognise and provide resources for the additional coordination work required to bridge disciplines so that collaboration is not slowed down by infrastructure and communication bottlenecks. Overall, this work provides insight into the complex navigation of early-stage AI development across different stakeholders in UK healthcare research consortia, including data scientists, clinicians, and those coordinating patient and public involvement. While direct interviews with patient and public contributors are still needed, our findings help establish new directions for research on how AI reshapes interdisciplinary collaboration in healthcare.

# Acknowledgements

We thank all the participants that volunteered to take part in our study and our colleagues that helped to shape the study by taking part in pilot interviews.

# References

1. Rajpurkar P, Chen E, Banerjee O, Topol EJ. AI in health and medicine. Nat Med. 2022 Jan;28(1):31–8. doi:10.1038/s41591-021-01614-0

2. Alowais SA, Alghamdi SS, Alsuhebany N, Alqahtani T, Alshaya AI, Almohareb SN, et al. Revolutionizing healthcare: the role of artificial intelligence in clinical practice. BMC Med Educ. 2023 Sep 22;23(1):689. doi:10.1186/s12909-023-04698-z

3. Yu KH, Beam AL, Kohane IS. Artificial intelligence in healthcare. Nat Biomed Eng. 2018 Oct;2(10):719–31. doi:10.1038/s41551-018-0305-z

4. Arencibia-Jorge R, Vega-Almeida RL, Jiménez-Andrade JL, Carrillo-Calvet H. Evolutionary stages and multidisciplinary nature of artificial intelligence research. Scientometrics. 2022 Sep 1;127(9):5139–58. doi:10.1007/s11192-022-04477-5

5. Van Noorden R. Interdisciplinary research by the numbers. Nature. 2015 Sep 1;525(7569):306–7. doi:10.1038/525306a

6. Hall KL, Vogel AL, Huang GC, Serrano KJ, Rice EL, Tsakraklides SP, et al. The science of team science: A review of the empirical evidence and research gaps on collaboration in science. Am Psychol. 2018;73(4):532–48. doi:10.1037/amp0000319

7. Whitfield K, Reid C. Assumptions, Ambiguities, and Possibilities in Interdisciplinary Population Health Research. Can J Public Health Rev Can Santee Publique. 2004;95(6):434–6.

8. Choi BCK, Pak AWP. Multidisciplinarity, interdisciplinarity and transdisciplinarity in health research, services, education and policy: 1. Definitions, objectives, and evidence of effectiveness. Clin Investig Med Med Clin Exp. 2006 Dec;29(6):351–64.

9. Smye SW, Frangi AF. Interdisciplinary research: shaping the healthcare of the future. Future Healthc J. 2021 Jul;8(2):e218–23. doi:10.7861/fhj.2021-0025

10. Elish MC, Watkins EA. Repairing Innovation: A Study of Integrating AI in Clinical Care [Internet]. Data & Society; 2020 Sep. Available from: https://datasociety.net/library/repairing-innovation/

11. Sartori L, Theodorou A. A sociotechnical perspective for the future of AI: narratives, inequalities, and human control. Ethics Inf Technol. 2022 Mar;24(1):4. doi:10.1007/s10676-022-09624-3

12. Winter P, Carusi A. 'If You're Going to Trust the Machine, Then That Trust Has Got to Be Based on Something':: Validation and the Co-Constitution of Trust in Developing Artificial Intelligence (AI) for the Early Diagnosis of Pulmonary Hypertension (PH). Sci Technol Stud. 2022 Mar 22. doi:10.23987/sts.102198

13. Bijker WE, Law J, editors. Shaping technology/building society: studies in sociotechnical change. Nachdr. Cambridge, Mass.: MIT Press; 2010. 341 p. (Inside technology).

14. Suchman L. The uncontroversial 'thingness' of AI. Big Data Soc. 2023 Jul;10(2):20539517231206794. doi:10.1177/20539517231206794

15. Lorenzini G, Arbelaez Ossa L, Milford S, Elger BS, Shaw DM, De Clercq E. The "Magical Theory" of AI in Medicine: Thematic Narrative Analysis. JMIR AI. 2024 Aug 19;3:e49795. doi:10.2196/49795

16. Piorkowski D, Park S, Wang AY, Wang D, Muller M, Portnoy F. How AI Developers Overcome Communication Challenges in a Multidisciplinary Team: A Case Study. Proc ACM Hum-Comput Interact. 2021 Apr 13;5(CSCW1):1–25. doi:10.1145/3449205

17. Pang RY, Wang R, Nelson J, Battle L. How Do Data Science Workers Communicate Intermediate Results? In: 2022 IEEE Visualization in Data Science (VDS) [Internet]. Oklahoma City, OK, USA: IEEE; 2022 [cited 2024 Feb 22]. p. 46–54. Available from: https://ieeexplore.ieee.org/document/9982308/ doi:10.1109/VDS57266.2022.00010

18. Roy A, Raghunandan D, Elmqvist N, Battle L. How I Met Your Data Science Team: A Tale of Effective Communication. In: 2023 IEEE Symposium on Visual Languages and Human-Centric Computing (VL/HCC) [Internet]. Washington, DC, USA: IEEE; 2023 [cited 2023 Dec 5]. p. 199–208. Available from: https://ieeexplore.ieee.org/document/10305667/ doi:10.1109/VL-HCC57772.2023.00032

19. Zhang AX, Muller M, Wang D. How do Data Science Workers Collaborate? Roles, Workflows, and Tools. Proc ACM Hum-Comput Interact. 2020 May 28;4(CSCW1):1–23. doi:10.1145/3392826

20. Mao Y, Wang D, Muller M, Varshney KR, Baldini I, Dugan C, et al. How Data Scientists Work Together With Domain Experts in Scientific Collaborations: To Find The Right

Answer Or To Ask The Right Question? Proc ACM Hum-Comput Interact. 2019 Dec 5;3(GROUP):1–23. doi:10.1145/3361118

21. Ross S, Lavis J, Rodriguez C, Woodside J, Denis JL. Partnership experiences: Involving decision-makers in the research process. J Health Serv Res Policy. 2003 Oct;8(2_suppl):26–34. doi:10.1258/135581903322405144

22. Sendak MP, Ratliff W, Sarro D, Alderton E, Futoma J, Gao M, et al. Real-World Integration of a Sepsis Deep Learning Technology Into Routine Clinical Care: Implementation Study. JMIR Med Inform. 2020 Jul 15;8(7):e15182. doi:10.2196/15182

23. Verma H, Mlynar J, Schaer R, Reichenbach J, Jreige M, Prior J, et al. Rethinking the Role of AI with Physicians in Oncology: Revealing Perspectives from Clinical and Research Workflows. In: Proceedings of the 2023 CHI Conference on Human Factors in Computing Systems [Internet]. Hamburg Germany: ACM; 2023 [cited 2023 Jun 27]. p. 1–19. Available from: https://dl.acm.org/doi/10.1145/3544548.3581506 doi:10.1145/3544548.3581506

24. Vijlbrief D, Dudink J, Van Solinge W, Benders M, Haitjema S. From computer to bedside, involving neonatologists in artificial intelligence models for neonatal medicine. Pediatr Res. 2023 Jan;93(2):437–9. doi:10.1038/s41390-022-02413-0

25. NHS England. NHS artificial intelligence (AI) giving patients better care and support [Internet]. 2024 Dec 12 [cited 2025 Jun 24]. Available from: https://www.england.nhs.uk/2024/12/nhs-ai-giving-patients-better-care-and-support/

26. Nilsen P, Svedberg P, Neher M, Nair M, Larsson I, Petersson L, et al. A Framework to Guide Implementation of AI in Health Care: Protocol for a Cocreation Research Project. JMIR Res Protoc. 2023 Nov 8;12:e50216. doi:10.2196/50216 PubMed PMID: 37938896; PubMed Central PMCID: PMC10666006.

27. Bastian G, Baker GH, Limon A. Bridging the divide between data scientists and clinicians. Intell-Based Med. 2022;6:100066. doi:10.1016/j.ibmed.2022.100066

28. Hogg HDJ, Al-Zubaidy M, Technology Enhanced Macular Services Study Reference Group, Talks J, Denniston AK, Kelly CJ, et al. Stakeholder Perspectives of Clinical Artificial Intelligence Implementation: Systematic Review of Qualitative Evidence. J Med Internet Res. 2023 Jan 10;25:e39742. doi:10.2196/39742 PubMed PMID: 36626192; PubMed Central PMCID: PMC9875023.

29. Fitzpatrick G, Ellingsen G. A Review of 25 Years of CSCW Research in Healthcare: Contributions, Challenges and Future Agendas. Comput Support Coop Work CSCW. 2013 Aug;22(4–6):609–65. doi:10.1007/s10606-012-9168-0

30. Lipton ZC. The mythos of model interpretability. Commun ACM. 2018 Sep 26;61(10):36–43. doi:10.1145/3233231

31. The Alan Turing Institute. AI for multiple long-term conditions: Research Support Facility [Internet]. 2023 [cited 2023 Dec 19]. Available from: https://www.turing.ac.uk/research/research-projects/ai-multiple-long-term-conditions-research-support-facility

32. consortium [Internet]. 2023 [cited 2023 Dec 20]. Available from: https://dictionary.cambridge.org/us/dictionary/english/consortium

33. NIHR. Artificial Intelligence for Multiple Long-Term Conditions (AIM) - Research Specification [Internet]. [cited 2023 Aug 8]. Available from: https://www.nihr.ac.uk/documents/artificial-intelligence-for-multiple-long-term-conditions-aim-research-specification/24646

34. NHS Transformation Directorate [Internet]. 2023 [cited 2024 Sep 30]. Artificial Intelligence. Available from: https://transform.england.nhs.uk/information-governance/guidance/artificial-intelligence/

35. Walker LE, Abuzour AS, Bollegala D, Clegg A, Gabbay M, Griffiths A, et al. The DynAIRx Project Protocol: Artificial Intelligence for dynamic prescribing optimisation and care integration in multimorbidity. J Multimorb Comorbidity. 2022;12:26335565221145493. doi:10.1177/26335565221145493 PubMed PMID: 36545235; PubMed Central PMCID: PMC9761229.

36. Malterud K, Siersma VD, Guassora AD. Sample Size in Qualitative Interview Studies: Guided by Information Power. Qual Health Res. 2016 Nov;26(13):1753–60. doi:10.1177/1049732315617444

37. Otter.ai. Otter.ai - Voice Meeting Notes & Real-time Transcription [Internet]. 2003 [cited 2023 Aug 8]. Available from: https://otter.ai/

38. Braun V, Clarke V. Using thematic analysis in psychology. Qual Res Psychol. 2006 Jan 1;3(2):77–101. doi:10.1191/1478088706qp063oa

39. Taguette. Welcome | Taguette [Internet]. [cited 2023 Aug 8]. Available from: https://app.taguette.org/

40. Moor J. The Dartmouth College Artificial Intelligence Conference: The Next Fifty Years. AI Mag. 2006;27(4):87. doi:https://doi.org/10.1609/aimag.v27i4.1911

41. Vrettas G, Sanderson M. Conferences versus journals in computer science. J Assoc Inf Sci Technol. 2015 Dec;66(12):2674–84. doi:10.1002/asi.23349

42. Pink S, Ruckenstein M, Willim R, Duque M. Broken data: Conceptualising data in an emerging world. Big Data Soc. 2018 Jan;5(1):2053951717753228. doi:10.1177/2053951717753228

43. Tanweer A, Fiore-Gartland B, Aragon C. Impediment to insight to innovation: understanding data assemblages through the breakdown–repair process. Inf Commun Soc. 2016 Jun 2;19(6):736–52. doi:10.1080/1369118X.2016.1153125

44. Lupton D. How do data come to matter? Living and becoming with personal data. Big Data Soc. 2018 Jul;5(2):2053951718786314. doi:10.1177/2053951718786314

45. Agrawal R, Prabakaran S. Big data in digital healthcare: lessons learnt and recommendations for general practice. Heredity. 2020 Apr;124(4):525–34. doi:10.1038/s41437-020-0303-2

46. Kim M, Zimmermann T, DeLine R, Begel A. The emerging role of data scientists on software development teams. In: Proceedings of the 38th International Conference on Software Engineering [Internet]. Austin Texas: ACM; 2016 [cited 2023 Jul 7]. p. 96–107. Available from: https://dl.acm.org/doi/10.1145/2884781.2884783 doi:10.1145/2884781.2884783

47. Kross S, Guo P. Orienting, Framing, Bridging, Magic, and Counseling: How Data Scientists Navigate the Outer Loop of Client Collaborations in Industry and Academia. Proc ACM Hum-Comput Interact. 2021 Oct 13;5(CSCW2):1–28. doi:10.1145/3476052

48. Stokols D, Misra S, Moser RP, Hall KL, Taylor BK. The ecology of team science: understanding contextual influences on transdisciplinary collaboration. Am J Prev Med. 2008 Aug;35(2 Suppl):S96-115. doi:10.1016/j.amepre.2008.05.003 PubMed PMID: 18619410.

49. Donia J, Shaw JA. Co-design and ethical artificial intelligence for health: An agenda for critical research and practice. Big Data Soc. 2021 Jul 1;8(2):20539517211065248. doi:10.1177/20539517211065248

50. Birhane A, Isaac W, Prabhakaran V, Diaz M, Elish MC, Gabriel I, et al. Power to the People? Opportunities and Challenges for Participatory AI. In: Equity and Access in Algorithms, Mechanisms, and Optimization [Internet]. Arlington VA USA: ACM; 2022 [cited 2023 Jul 11]. p. 1–8. Available from: https://dl.acm.org/doi/10.1145/3551624.3555290 doi:10.1145/3551624.3555290

51. Thompson A, Bartle V, Remfry EA, Reynolds DJ, Barnes MR, Reynolds NJ, et al. Public and Patient Involvement in Artificial Intelligence and Big Data Healthcare Research: An Exploration of Issues and Challenges Within the AI-Multiply Project. Health Expect. 2025 Dec;28(6):e70490. doi:10.1111/hex.70490

52. Remfry E, Reynolds DJ, De Queiroz SM, Social Action For Health, Mathur R, Barnes MR, et al. Using Arts-Based Methods to Involve People Living in Tower Hamlets With Multiple Long-Term Conditions in the Development of Artificial Intelligence Tools in Healthcare Research. Health Expect. 2026 Apr;29(2):e70621. doi:10.1111/hex.70621

53. Reynolds DJ, Mountain S, Bartle V, Remfry E, Barnes MR, Reynolds NJ, et al. Targeting everyday decision makers in research: early career researcher and patient and public involvement and engagement collaboration in an AI-in-healthcare project. Res Involv Engagem. 2025 Aug 19;11(1):100. doi:10.1186/s40900-025-00753-9

54. Banerjee S, Alsop P, Jones L, Cardinal RN. Patient and public involvement to build trust in artificial intelligence: A framework, tools, and case studies. Patterns. 2022 Jun 10;3(6):100506. doi:10.1016/j.patter.2022.100506 PubMed PMID: 35755870; PubMed Central PMCID: PMC9214308.

55. Nyström ME, Karltun J, Keller C, Andersson Gäre B. Collaborative and partnership research for improvement of health and social services: researcher’s experiences from 20 projects. Health Res Policy Syst. 2018 May 30;16(1):46. doi:10.1186/s12961-018-0322-0

56. Lomas J. The in-between world of knowledge brokering. BMJ. 2007 Jan 18;334(7585):129–32. doi:10.1136/bmj.39038.593380.AE PubMed PMID: 17235094.

57. Galsgaard A, Doorschodt T, Holten AL, Müller FC, Ploug Boesen M, Maas M. Artificial intelligence and multidisciplinary team meetings; a communication challenge for radiologists’ sense of agency and position as spider in a web? Eur J Radiol. 2022 Oct;155:110231. doi:10.1016/j.ejrad.2022.110231

58. Groves L, Peppin A, Strait A, Brennan J. Going public: the role of public participation approaches in commercial AI labs [Internet]. arXiv; 2023 [cited 2023 Aug 25]. Available from: http://arxiv.org/abs/2306.09871 doi:10.48550/arXiv.2306.09871

59. UKRI. Shared commitment to improve public involvement in research [Internet]. 2022 [cited 2023 Aug 25]. Available from: https://www.ukri.org/news/shared-commitment-to-improve-public-involvement-in-research/

60. Alberto IRI, Alberto NRI, Ghosh AK, Jain B, Jayakumar S, Martinez-Martin N, et al. The impact of commercial health datasets on medical research and health-care algorithms. Lancet Digit Health. 2023 May;5(5):e288–94. doi:10.1016/S2589-7500(23)00025-0

61. Essex R, Kennedy J, Miller D, Jameson J. A scoping review exploring the impact and negotiation of hierarchy in healthcare organisations. Nurs Inq. 2023;n/a(n/a):e12571. doi:10.1111/nin.12571

62. Croxford L, Raffe D. The iron law of hierarchy? Institutional differentiation in UK higher education. Stud High Educ. 2015 Oct 21;40(9):1625–40. doi:10.1080/03075079.2014.899342